\documentclass[aps,prl,showpacs,twocolumn,superscriptaddress,epsf]{revtex4}
\usepackage{graphicx}
\usepackage{psfig}
\usepackage{epsfig}

\begin{document}

\title{ Isospin Dependence of $^1S_0$ Proton and Neutron Superfluidity in Asymmetric Nuclear Matter}

 \author{W. Zuo$^1$\footnote{Corresponding address:
          Institute of Modern Physics, Chinese Academy of Science,
          P.O. Box 31, Lanzhou 730000, P.R. China.
          Tel: 0086-931-4969318; E-mail: zuowei@impcas.ac.cn}, and G. C. Lu}
\affiliation{Institute of Modern Physics, Chinese Academy of
Science, Lanzhou 730000, China}

\date{\today}
\begin{abstract}
We investigate the $^1S_0$ neutron and proton superfluidity in
isospin asymmetric nuclear matter. We have concentrated on the
isospin dependence of the pairing gaps and the effect of a
microscopic three-body force. It is found that as the isospin
asymmetry goes higher, the neutron $^1S_0$ superfluid phase
shrinks gradually to a smaller density domain, while the proton
one extends rapidly to a much wider density domain. The three-body
force turns out to weaken the neutron $^1S_0$ superfluidity
slightly, but it suppresses strongly the proton $^1S_0$
superfluidity at high densities in nuclear matter with large
isospin asymmetry.
\end{abstract}

\maketitle

 Superfluidity plays an important role in understanding
a number of astrophysical phenomena in neutron
stars~\cite{dean:2003,lombardo:2001r, page:1994,heiselberg:2000,
link:2003,shapiro:1983,sauls:1989,pines:1985,gusakov:2005}. In the
inner crust of a neutron star
 where the total
 baryon density $\rho$ is low, neutrons are thought to be
 superfluid in the $^1S_0$ channel. In the nuclear core
 part (i.e, the outer core part),
 protons may form a superfluid in the $^1S_0$ partial wave states,
 and neutron superfluidity is expected in the $^3P_2$-$^3F_2$
 partial wave channel.
It is generally expected that the cooling processes via neutrino
emission~\cite{page:1994,heiselberg:2000,link:2003}, the
properties of rotating dynamics, the post-glitch timing
observations~\cite{shapiro:1983,sauls:1989}, the possible vertex
pinning~\cite{pines:1985} of a neutron star are rather sensitive
to the presence of neutron and proton superfluid phases as well as
to their pairing strength. Recently it has been
shown~\cite{gusakov:2005} that a weak proton $^1S_0$ superfluidity
is in consistent with the cooling data if the effect of the
possible presence of accreted envelopes is taken into account in
the cooling scenario.

 Nucleon superfluidity in symmetric nuclear matter and pure
 neutron matter has been investigated extensively by many
 authors using various theoretical
aproaches\cite{baldo:1990,elgaroy:1996a,duguet:2004,
fabrocini:2005,clark:1976,
anisworth:1989,pethick:1991,schulze:1996,
serra:2001,rabhi:2002,schwenk:2003,muther:2005,baldo:2000,
bozek:2000,zuo:2002c}. All these investigations have predicted the
occurrence of the $^1S_0$ nucleon superfluid phase in the low
density region although the obtained strengthes of the
superfluidity are remarkably sensitive to the different approaches
and the different approximations adopted. There also exists few
investigations on the nucleon superfluidity in $\beta$-stable
neutron star matter~\cite{baldo:1992}. It is shown that due to the
small proton fraction in $\beta$-stable matter, the proton
superfluid phase in the $^1S_0$ channel may extend to much higher
baryon densities with a much smaller maximum of the pairing gap as
compared to the case of symmetric nuclear matter. Therefore it is
of interest to investigate the variation of nucleon superfluidity
vs. isospin asymmetry in asymmetric nuclear matter, which is
expected to be helpful for understanding the properties of nucleon
superfluid phases in neutron stars.

 The aim of this work is
devoted to the isospin dependence of the $^1S_0$ neutron and
proton superfluid phases in asymmetric nuclear matter, and to
investigating the influence of three-body forces which turn out to
be crucial for reproducing the empirical saturation properties of
nuclear matter in a non-relativistic microscopic
approach~\cite{grange:1989,zuo:2002a,mac}.

For such a purpose, we shall not go beyond the BCS framework. In
this case, the pairing gap which characterizes the superfluidity
in a homogeneous Fermi system is determined by the standard BCS
gap equation~\cite{RING}, i.e.,
\begin{eqnarray}\label{e:gap}
 \Delta_{\vec{k}} = -\sum_{\vec{k'}}
 v(\vec{k},\vec{k}') \frac{1}{2E_{\vec{k}'}}
 {\Delta_{\vec{k}'}} \:,
\end{eqnarray}
where $v(\vec{k},\vec{k}')$ is the realistic $NN$ interaction in
momentum space, $E_{\vec{k}}=\sqrt{(\epsilon_{\vec{k}}
-\epsilon_F)^2+\Delta_{\vec{k}}^2}$, $\epsilon_{\vec{k}}$ and
$\epsilon_F$ being the s.p. energy and its value at the Fermi
surface, respectively. In the BCS gap equation, the most important
ingredients are the realistic $NN$ interaction
$v(\vec{k},\vec{k}')$ and the neutron and proton s.p. energies
$\epsilon_{\vec{k}}$ in asymmetric nuclear matter. For the $NN$
interaction, we adopt the Argonne $V18$ ($AV_{18}$) two-body
interaction~\cite{wiringa:1995} plus a microscopic three-body
force (TBF). The TBF adopted in the present calculation was
originally proposed in Ref.~\cite{grange:1989} based on the
meson-exchange current approach. The parameters of the TBF, i.e.,
the coupling constants and the form factors, were
determined~\cite{zuo:2002a} from the one-boson-exchange potential
model to meet the self-consistent requirement with the adopted
$AV_{18}$ two-body force. A more detailed description of the TBF
model and the related approximations can be found in
Ref.~\cite{grange:1989}.

The proton and neutron s.p. energies in asymmetric nuclear matter
are calculated by using the BHF approach for isospin asymmetric
nuclear matter~\cite{zuo:1999}. The starting point of the BHF
approach is the Brueckner $G$-matrix, which satisfies the
Bethe-Goldstone (BG) equation~\cite{jeukenne:1976}:
\begin{equation} G(\rho,\beta;\omega)
= v + v \sum_{k_1 k_2}
 \frac { |k_1 k_2 \rangle Q(k_1,k_2) \langle k_1 k_2| }
{\omega - \epsilon(k_1)-\epsilon(k_2)+i\eta} G(\rho,\beta;\omega),
\label{eq:BG}
\end{equation}
where $k_i\equiv(\vec k_i,\sigma_i,\tau_i)$ denotes the momentum,
the $z$-components of spin and isospin of a nucleon, respectively.
$\omega$ is the starting energy and $Q(k_1,k_2)$ is the Pauli
operator. The isospin asymmetry parameter is defined as
$\beta=(\rho_n-\rho_p)/\rho$, being $\rho_n$, $\rho_p$, and $\rho$
the neutron, proton and total nucleon number densities,
respectively. The s.p. energy is given by $\epsilon(k) =
\hbar^2k^2/(2m)  +  U(k)$. In solving the BG equation, we adopt
the continuous choice~\cite{jeukenne:1976} for the s.p. potential
$U(k)$ since it has been proved to provide a much faster
convergency than the gap choice~\cite{song:1998}. Under the
continuous choice, the s.p. potential describes physically the
nuclear mean field felt by a nucleon in nuclear
medium~\cite{lejeune:1978} and
 is calculated from the real part of the on-shell $G$-matrix.

The neutron and proton superfluidity in nuclear matter can be
described by their pairing energy gaps at their respective Fermi
surfaces. To solve the gap equation, we follow the scheme given in
Ref.~\cite{baldo:1990}. The proton and neutron s.p. energies in
the gap equation are calculated from the BHF approach. In our
calculations, the TBF contribution has been included by reducing
the TBF to an equivalently effective two-body interaction
according to the standard scheme as described in
Ref.~\cite{grange:1989}. A detailed description and justification
of the method are discussed in Refs.~\cite{grange:1989,zuo:2002a}.

In Fig.~\ref{f:neutrongap} is shown the neutron energy gap in the
$^1S_0$ partial wave channel $\Delta^n_F=\Delta(k^n_F)$ as a
function of the total baryon density $\rho$. The curves along the
direction of the arrow from the bottom correspond to $\beta$ =
0.2, 0.4, 0.6, and 0.8, respectively. We see that the neutron
$^1S_0$ superfluid phase exists only at low densities ($\rho\le
0.13$ fm$^{-3}$) and the peaks of the pairing gaps are located
around $\rho=0.02$ fm$^{-3}$, which is compatible with the
previous predictions for pure neutron matter and symmetric nuclear
matter~\cite{baldo:1990,zuo:2002c}. As the isospin asymmetry
$\beta$ increases, the neutron fraction increases for a given
total density and as a consequence the density domain for the
neutron $^1S_0$ superfluidity shrinks gradually. In the case
without including the TBF, the density region for the presence of
the neutron superfluidity reduces from $\rho\le 0.13$ fm$^{-3}$
down to $\rho\le0.1$ fm$^{-3}$ as the isospin asymmetry rises from
$\beta=0.2$ to $\beta=0.8$. In Ref.\cite{zuo:2002c}, we calculated
the $^1S_0$ pairing gaps in symmetric nuclear matter ($\beta=0$)
and pure neutron matter ($\beta=1$). We found that the neutron
$^1S_0$ superfluid phase is expected to appear in a density range
of $\rho_B\le 0.135$ fm$^{-3}$ for symmetric nuclear matter and
$\rho_B\le 0.09$ fm$^{-3}$ for pure neutron matter, in consistent
with our present results. The above obtained isospin dependence of
the pairing gap is readily understood as follows. In asymmetric
nuclear matter, the strength of the neutron pairing gap is related
directly with the {\it neutron} number density. As the asymmetry
rises, the neutron excess and the neutron number density for a
given total nucleon density increase. As a consequences, the
density domain for the existence of the neutron superfluid phase
shrinks gradually going from symmetric matter to pure neutron
matter. We notice that in both cases with and without including
the TBF, the peak values of the gaps decrease and their location
shift slightly to lower densities as the matter becomes more
isospin asymmetric. This is mainly attributed to the isospin
dependence of the neutron s.p. energy spectrum in asymmetric
nuclear matter. As we know, the neutron s.p. potential in
asymmetric matter becomes less deeper and the neutron effective
mass gets larger for a larger isospin asymmetry according to the
BHF and DBHF calculations~\cite{zuo:1999,dbhf}. We may also see
from Fig.~\ref{f:neutrongap} that within the region of the neutron
superfluid phase, as the isospin asymmetry increases, the pairing
gap becomes smaller at low densities below and in the vicinity of
the peak position, while it gets larger at relatively high
densities. The above obtained isospin behavior stems from the
competition between two different mechanisms. On the one hand, at
a fixed total density on the right side of the peak position, a
higher asymmetry corresponds to a larger value of neutron density
and a weaker neutron superfluidity. On the other hand, the isospin
dependence of the neutron s.p. spectrum intends to enhance the
neutron pairing gap as the isospin asymmetry increases. At low
densities around and below the peak position, the later effect is
dominated, while as the density increases, the former mechanism
becomes more and more effective. Comparing the solid curves with
the corresponding dashed ones, we may see that the TBF affects
mainly the pairing gap at high densities. Inclusion of the TBF
weakens the neutron superfluidity at high densities and makes the
predicted density range for the existence of neutron superfluid
phase smaller.

In Fig.~\ref{f:protongap} is reported the proton $^1S_0$ pairing
gap in asymmetric nuclear matter for four different asymmetries
$\beta=0.2$, 0.4, 0.6, and 0.8, respectively. The solid curves are
obtained by including the TBF, while the dashed ones without
including the TBF. The isospin dependence of the proton $^1S_0$
pairing gaps turns out to be completely different from that of the
corresponding neutron ones in both cases with and without the TBF.
It is seen that as the isospin asymmetry increases, the density
domains for the existence of the proton superfluid phases enlarge
rapidly, the peaks of the pairing gaps become lower appreciably
and shift to higher densities gradually. As the asymmetry rises
from $\beta=0.2$ to $\beta=0.8$, the density domain extends from
$\rho\le 0.17$ fm$^{-3}$ to $\rho\le 0.435$ fm$^{-3}$ in the case
of not including the TBF and from $\rho\le 0.13$ fm$^{-3}$ to
$\rho\le 0.25$ fm$^{-3}$ in the case of including the TBF, the
peak values lower from about 2.0 MeV to about 1.4 MeV in both
cases. The difference between the proton and neutron superfluid
phases is especially pronounced at high isospin asymmetries. As
compared to the neutron $^1S_0$ superfluidity, the proton $^1S_0$
superfluid phase in highly asymmetric matter extends to much
higher densities but with a smaller peak value of the pairing gap.
For example, in the case without including the TBF, at
$\beta=0.8$, the proton $^1S_0$ superfluidity is predicted to
exist in the density region of $\rho\le 0.435$ fm$^{-3}$ which is
much wider than that of $\rho\le 0.1$ fm$^{-3}$ for the neutron
$^1S_0$ superfluidity. The above behavior of the proton
superfluidity vs. isospin asymmetry can be readily explained in
terms of the isospin dependence of the proton density and the
proton s.p. spectrum in asymmetric nuclear matter. First, at a
fixed total density, a higher asymmetry corresponds to a smaller
proton concentration and a lower proton density. Such an isospin
dependence of the proton density is directly responsible for the
widening of the density domain as a function of asymmetry. Second,
the proton s.p. potential becomes deeper going from symmetric
matter to pure neutron matter~\cite{zuo:1999}, and this results in
the lowering of the proton pairing peak vs. asymmetry. At
relatively low densities below and around the peak position, the
variation of the proton s.p. spectrum as a function of asymmetry
plays an major role in determining the isospin variation of the
proton pairing gap and thus the proton superfluidity becomes
weaker at a higher asymmetry. While as the density increases, the
effect due to the decreasing of the proton fraction vs. asymmetry
gets stronger. At high enough densities, it becomes predominant
and leads to a turnover of the isospin behavior of the proton
superfluidity (i.e., a stronger superfluidity at a higher
asymmetry). The turnover is clearly seen in Fig.~\ref{f:protongap}
and stems from the competition between the two above-mentioned
isospin effects.

We notice that the TBF effect is negligibly small at low densities
below and around the peak position. However, it gets stronger
rapidly as the density goes up. The TBF turns out to induce a
significant reduction of the proton pairing gaps in the
high-density superfluidity domain. Consequently, it leads to a
remarkable shrinking of the density domain for the existence of
the proton superfluid phase. We notice that the above predicted
TBF suppression of the $^1S_0$ proton superfluidity is
particularly pronounced for highly asymmetric matter. For example,
at $\beta=0.8$, the density domain of the proton superfluid phase
is reduced by about $50\%$, i.e., from $\rho\le 0.435$ fm$^{-3}$
down to $\rho\le 0.25$ fm$^{-3}$, by inclusion of the TBF. In
nuclear matter, proton pairs are embedded in the medium of
neutrons and protons, both the surrounded protons and neutrons
contribute to the TBF renormalization of the proton-proton
interaction, therefore the TBF effect on the proton pairing
correlations is determined by the {\it total} nucleon number
density instead of the {\it proton} number density. Accordingly,
in spite of the small proton fractions and the low proton
densities at high asymmetries, the TBF modifies strongly the
proton-proton pairing interactions at high-density superfluidity
domain and weakens considerably the corresponding proton pairing
gaps. One can verify readily from Fig.~\ref{f:protongap} that the
TBF suppression of the proton superfluidity is mainly in the
high-density region and the reduction of the gap increases rapidly
as increasing the total density. We also notice from
Fig.~\ref{f:protongap} that the TBF effect gets stronger as the
asymmetry increases since the proton superfluid phase extends to
larger densities for asymmetric matter at higher isospin
asymmetries. Inclusion of the TBF weakens considerably the isospin
dependence of the predicted proton superfluidity as compared to
the results obtained by adopting purely the $AV_{18}$ two-body
force.

In summary, we have calculated the neutron and proton $^1S_0$
pairing gaps in asymmetric nuclear matter based on the BHF
approach and the BCS theory. We have especially investigated the
isospin dependence of the pairing gaps in the $^1S_0$ channel and
the influence of the TBF.  It is shown that the isospin dependence
of the proton $^1S_0$ superfluidity in asymmetric nuclear matter
is completely different from that of the neutron one. The neutron
$^1S_0$ superfluid phase exists only in low density region for all
isospin asymmetries. As the matter goes from symmetric nuclear
matter to pure neutron matter, the peak value of the neuron
$^1S_0$ pairing gap becomes larger and the density domain for the
existence of neutron superfluidity shrinks gradually. While the
density domain for the proton superfluid phase enlarges rapidly as
the isospin asymmetry rises and it may extend to very high
densities for highly asymmetric nuclear matter. The peak value of
the proton pairing gap turns out to be larger in asymmetric
nuclear matter at a higher isospin asymmetry.

The TBF affects only weakly the neutron $^1S_0$ superfluid phase
in asymmetric nuclear matter, i.e., it reduces slightly the
pairing gap, due to the low-density region for this kind of
superfluidity. However it suppresses strongly the proton
superfluidity in the $^1S_0$ channel at high densities, especially
at high asymmetries. The density domain for the existence of the
proton $^1S_0$ superfluid phase is reduced by about $50\%$ from
$\rho\le 0.435$ fm$^{-3}$ to $\rho\le 0.25$ fm$^{-3}$ by inclusion
of the TBF.

The work is supported in part by the National Natural Science
Foundation of China (10575119), the Asia-Link project
(CN/ASIA-LINK/008(94791)) of the European Commission, and DFG,
Germany.

\noindent
 {\bf Figure Captions:}
\begin{figure}[ht]
\caption{ Neutron $^1S_0$ energy gap as a function of density in
asymmetric nuclear matter at various asymmetries. The solid curves
are predicted by adopting purely the $AV_{18}$ two-body
interaction, while the dashed ones by using the $AV_{18}$ plus the
TBF. } \label{f:neutrongap}
\end{figure}

\begin{figure}[ht]
\caption{The same as Fig.~\ref{f:neutrongap} but for proton
$^1S_0$ pairing gap.} \label{f:protongap}
\end{figure}

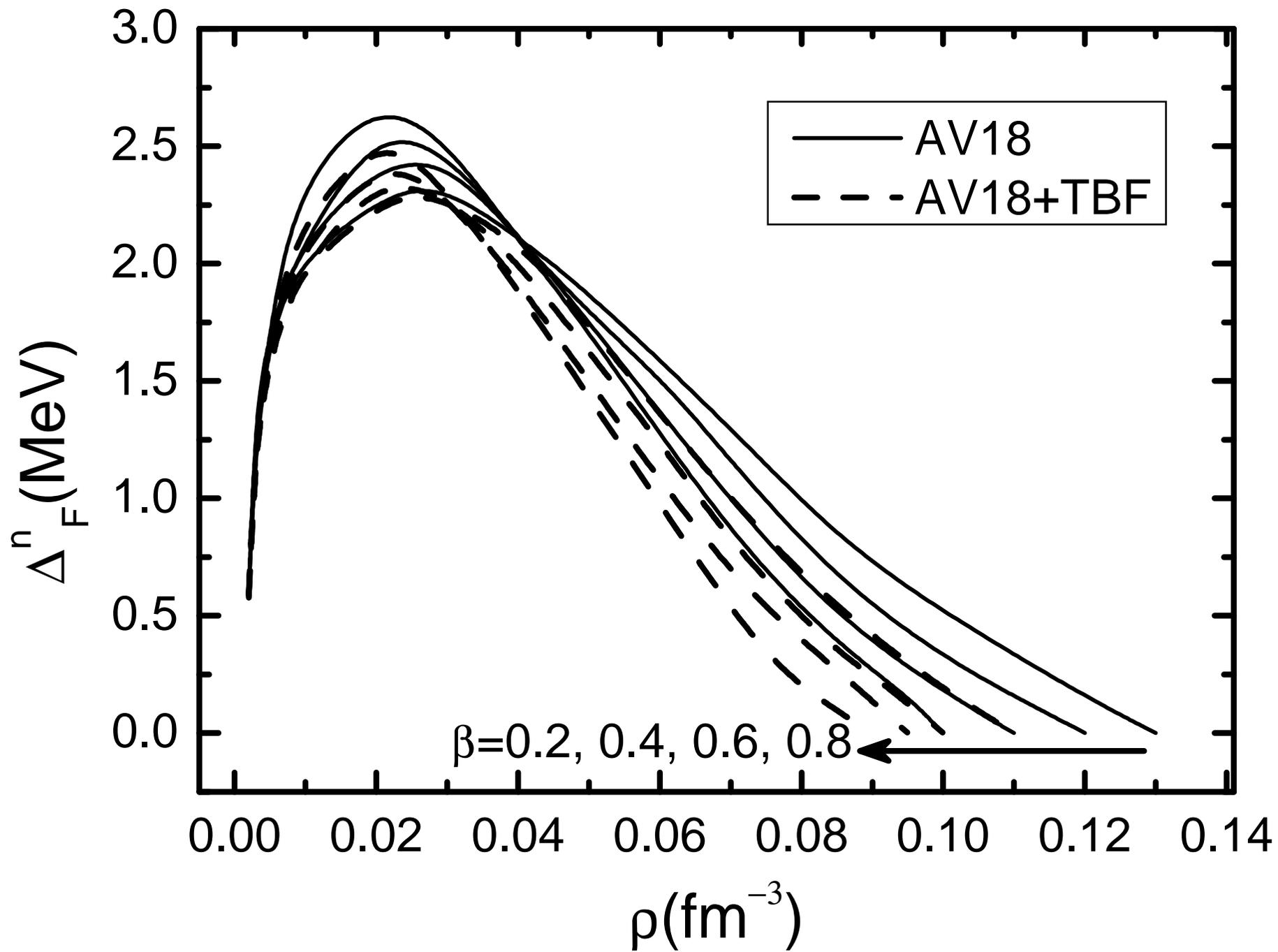
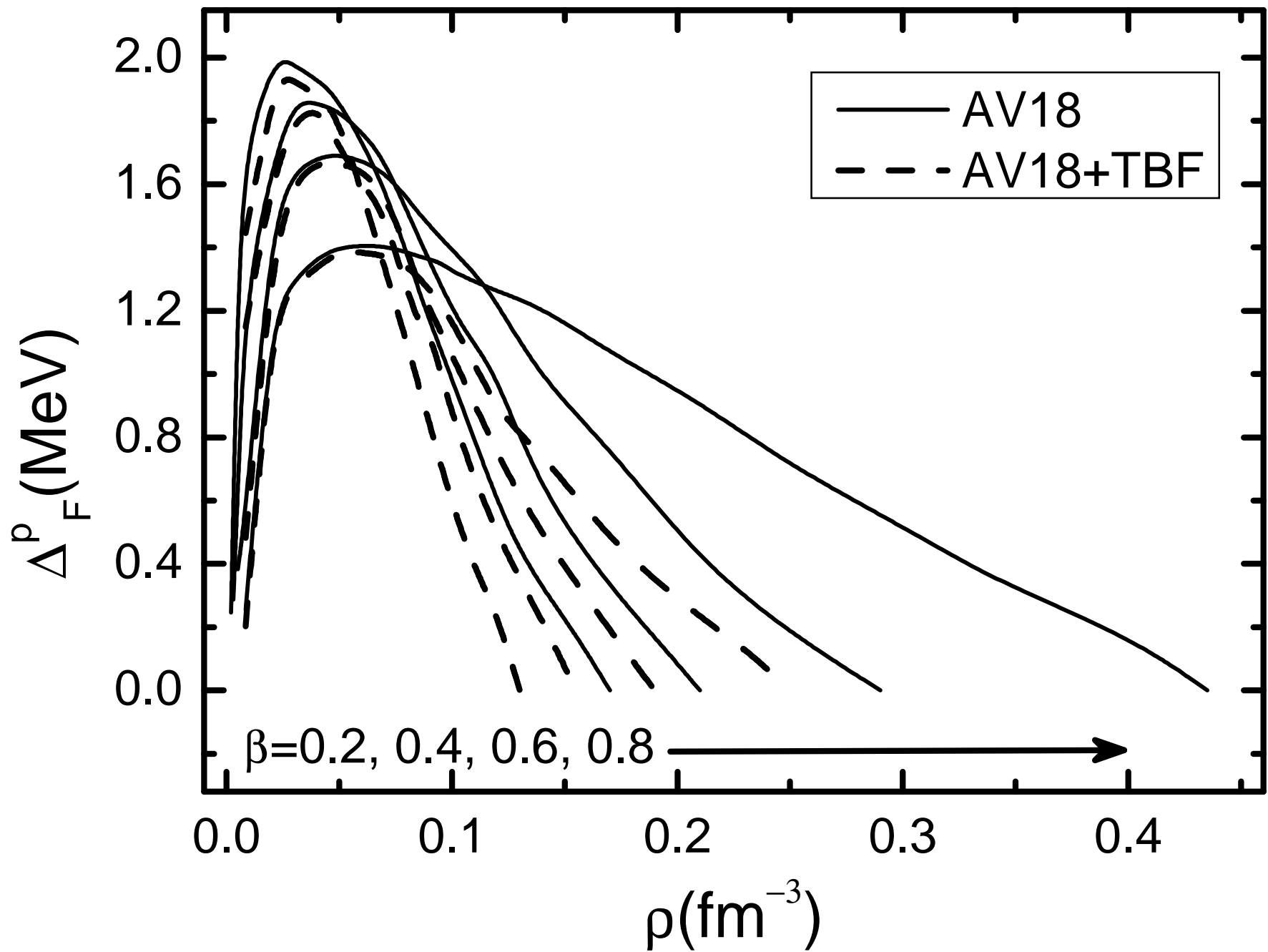
\end{document}